\documentclass[a4paper,11pt]{article}
\usepackage{jcappub} 

\usepackage[T1]{fontenc} 

\title{\boldmath Generalized Logotropic Models and their Cosmological Constraints}

\author[a,1]{H.B. Benaoum,\note{Corresponding author.}}
\author[b]{Pierre-Henri Chavanis,}
\author[c,d,e]{Hernando Quevedo}

\affiliation[a]{Department of Applied Physics and Astronomy,University of Sharjah, United Arab Emirates}
\affiliation[b]{Laboratoire de Physique Th\'eorique, Universit\'e de Toulouse, CNRS, UPS,France}
\affiliation[c]{Instituto de Ciencias Nucleares, Universidad Nacional Aut\'onoma deM\'exico, AP 70543, Ciudad de M\'exico 04510, Mexico}
\affiliation[d]{Dipartimento di Fisica and ICRANet, Universit\`a di Roma ``LaSapienza",  I-00185 Roma, Italy}
\affiliation[e]{Institute of Experimental and Theoretical Physics, Al-Farabi KazakhNational University, Almaty 050040, Kazakhstan}

\emailAdd{hbenaoum@sharjah.ac.ae}
\emailAdd{chavanis@irsamc.ups-tlse.fr}
\emailAdd{quevedo@nucleares.unam.mx}

\abstract{
We propose a new class of cosmological unified dark sector models called ``{\em
Generalized Logotropic Models}". They depend on a free parameter $n$.
The original logotropic model [P.H. Chavanis, Eur. Phys. J. Plus {\bf 130}, 130
(2015)] is a special case of our
generalized model corresponding to $n=1$. In our
scenario, the Universe is filled with a single fluid, a generalized
logotropic dark fluid (GLDF), whose pressure $P$ includes higher order
logarithmic terms of the rest-mass density $\rho_m$. The total
energy density $\epsilon$ is the sum of the rest-mass energy density $\rho_m
c^2$ and the internal energy density $u$ which play the role of dark matter
energy density $\epsilon_m$ and dark energy density $\epsilon_{de}$,
respectively. We investigate the cosmological behavior of the
generalized logotropic models by focusing on the evolution of the energy
density, scale factor, equation of state parameter, decceleration parameter and
squared speed of
sound. Low values of $n\le 3$ are favored.  We also study the
asymptotic behavior of the generalized logotropic models.  In particular, we 
show that the model 
presents a phantom behavior and has three distinct ways of evolution depending
on the value
of $n$. For $n\le 2$, it leads to a little rip and for $n>2$ to
a big rip. We predict the value of the big rip time as a function of $n$
without any free (undetermined) parameter.}

\keywords{dark energy theory, dark matter theory, cosmology, unified dark sector, logotropic model}

\begin{document}
\maketitle
\flushbottom

\section{Introduction}

Cosmological observations from various independent research teams show that the
current Universe is accelerating
\cite{riess1998,perlmutter1999,ade2014,spergel2003,tegmark}. This
accelerating expansion is due to an unknown component called ``dark energy"
which
works against gravity. The nature of the dark sector is still unknown and
several alternative models of dark matter (DM) and dark energy (DE) have been
proposed to account for the observation of the present cosmic acceleration. The
standard cold dark matter ($\Lambda$CDM) model is the simplest DE model and
relies on a cosmological constant to drive the current acceleration of the
Universe and on the existence of a pressureless DM to explain the
observed properties of the large-scale structures of the Universe
\cite{sahni2000,amendola2013}. However, the $\Lambda$CDM model
suffers from the cosmological constant problem
\cite{weinbergcosmo,paddycosmo}, namely why the  value of the cosmological
constant is so tiny, and the cosmic coincidence problem \cite{stein1,zws},
namely  why DM and DE are of similar
magnitudes today although they scale differently with the universe's
expansion. The CDM model also faces important
problems at the scale of DM halos such as the core-cusp problem
\cite{moore}, the missing satellite problem
\cite{satellites1,satellites2,satellites3}, and the ``too big to fail'' problem
\cite{boylan}. This leads to the so-called small-scale crisis of CDM 
\cite{crisis}. Among the wide range of
alternative DE models that have been proposed in the literature
(quintessence, k-essence, Chaplygin gas, tachyons,
phantom fields, holography...), an
interesting
class of dynamical DE
models considers DM and DE as different manifestations of a single-component
underlying fluid, often assumed to be a perfect fluid. The unified dark matter
and dark energy (UDM) models have the remarkable feature of describing the
dark sector of the
Universe as a single component which behaves as DM at early times and as DE at
late times
\cite{kamenshchik2001,benaoum2002,benaoum2012,benaoum2014,benaoum2019,dbc,
bento2002,bento2004,gorini2003,chavanis2015,chavanis2016,chavanis2018,ferreira,
boshkayev2019,luongo2019,capozziello2018,cosmopoly1,cosmopoly2,cosmopoly3,
universe} . The
cosmological
aspects of these UDM models have been studied recently in refs.
\cite{boshkayev2019,luongo2019,capozziello2018,li2019,chavanis2017,mamon2020,boshkayev2021testing}.
In particular, the logotropic model is
a candidate for unifying DE and DM \cite{chavanis2015}.
The logotropic model is able to
account for the
transition between a DM era and a DE era and is indistinguishable from the
$\Lambda$CDM model, for what concerns the evolution of the cosmological
background, up to $25$ billion years in the future  when it becomes phantom
\cite{chavanis2015,chavanis2016,chavanis2017,chavanis2018}. Remarkably,
the logotropic model implies
that DM halos should have a
constant surface density  and it predicts its universal value $\Sigma_0^{\rm
th}=0.01955  c\sqrt{\Lambda}/G=133\, M_{\odot}/{\rm pc}^2$
\cite{chavanis2015,chavanis2016,chavanis2017,chavanis2018} without
adjustable parameter. This
theoretical value is in good agreement with the value $\Sigma_0^{\rm
obs}=141_{-52}^{+83}\, M_{\odot}/{\rm pc}^2$ obtained from the observations
\cite{donato}. The logotropic model also predicts the values of
the present proportion of DM and DE
$\Omega_{dm,0}=\frac{1}{1+e}(1-\Omega_{b,0})=0.256$ and
$\Omega_{de,0}=\frac{e}{1+e}(1-\Omega_{b,0})=0.695$  \cite{chavanis2018}, in
good agreement with the observed values
$\Omega_{dm,0}=0.259$,
$\Omega_{de,0}=0.691$ and $\Omega_{b,0}=0.0486$.

In this work, we introduce a new class of UDM models called ``Generalized
Logotropic Models"  characterized by a single fluid equation of state (EoS) of
the form
\begin{eqnarray}
P  =   \sum_{i=0}^{N} A_i \ln^i \left(\frac{\rho_m}{\rho_{P}} \right),
\label{eq1}
\end{eqnarray}
where $P$ is the pressure, $\rho_m$ is the rest-mass density,  $A_i$ are
constants with the dimension of an energy density and
$\rho_{P}$ is a constant with the dimension of a mass density.
The above EoS returns the standard $\Lambda$CDM model for $N=0$ and the original logotropic model \cite{chavanis2015} for $N=1$. Following
\cite{chavanis2015}, we shall identify $\rho_P$ with the Planck
density $\rho_{P}={c^5}/({\hbar
G^2})=5.16\times 10^{99}\, {\rm g\, m^{-3}}$. The
single fluid which obeys Eq. (\ref{eq1}) will be
called the generalized logotropic dark fluid (GLDF). 

In this paper, we are interested in studying the dynamical evolution of various
generalized logotropic models. In particular, we examine in detail various
forms of generalized logotropic EoS and investigate how the Universe evolution 
is affected by the corresponding EoS. 
In section
\ref{sec:genlog}, we provide an
approach to motivate the logotropic model.
In section \ref{sec:cosm}, we consider an
isotropic,
homogeneous and spatially flat Universe  and introduce the main
equations characterizing generalized logotropic models. In section
\ref{sec_part}, we consider a subclass of  models with
$A_i=A_{n}\delta_{i,n}$ that are more easily tractable. In section
\ref{sec_evo}, we investigate the cosmological
behavior of the generalized logotropic models and focus on the evolution
of the  energy density,
scale factor, EoS parameter, decceleration parameter and squared
speed of sound. We also derive analytically the asymptotic
behavior of the
generalized logotropic models and distinguish three types of evolution
depending on the value of $n$. Finally,
section \ref{sec_con} is devoted to remarks and conclusions.
In Appendix \ref{sec_ae}, we show that the GLDF is
asymptotically equivalent to a form of Modified Chaplygin Gas (MCG). In
Appendix \ref{sec_intro}, we consider the two-fluid model associated with the
GLDF and determine the EoS of DE. In Appendix \ref{sec_arg},
we derive the present proportion of DM and DE by taking into account the
presence of baryons.

\section{Generalized Logotropic Equation of State}
\label{sec:genlog}

We consider the scenario where both DM and DE originate from a
single dark fluid. We first recall the justification of the standard logotropic
equation of state given in  \cite{chavanis2015}. In the Newtonian regime, the
condition of hydrostatic
equilibrium, which describes the balance between the gravitational
force and the pressure gradient, is given by
\begin{eqnarray}
\nabla P + \rho \nabla \Phi  = {\bf 0}.
\label{eq2}
\end{eqnarray}
Here $P$ and $\rho$  denote the pressure and  mass density of the fluid,
respectively. Moreover,  $\Phi$ is the gravitational potential. We assume one of
the simplest direct relations between the mass density $\rho$ and the pressure
$P$ of the fluid given by the polytropic EoS
\begin{eqnarray}
P  =  K \rho^{\gamma},
\label{eq3}
\end{eqnarray}
where $\gamma$ is the polytropic index and $K$ is a constant. Using the above
equation, the condition of hydrostatic equilibrium becomes
\begin{eqnarray}
K \gamma \rho^{\gamma -1} \nabla \rho + \rho \nabla \Phi  =  {\bf 0}.
\label{eq4}
\end{eqnarray}
Considering the limit $\gamma\rightarrow 0$, $K\rightarrow +\infty$
with $A=K\gamma$ fixed \cite{chavanis2015}, we get
\begin{eqnarray}
A\frac{\nabla \rho}{\rho} + \rho \nabla \Phi  = {\bf 0}.
\label{eq5}
\end{eqnarray}
By comparing equations (\ref{eq5}) and (\ref{eq2}), we find that the EoS
of the logotropic gas reads \cite{chavanis2015}
\begin{eqnarray}
P  =  A\ln\left (\frac{\rho}{\rho_*}\right ),
\label{eq8}
\end{eqnarray}
where $\rho_*$ is an integration constant. In this paper, we consider a
generalization of this EoS of the form
\begin{eqnarray}
P  =   \sum_{i=0}^{N} A_i \ln^i \left(\frac{\rho}{\rho_{*}} \right),
\label{eq1gr}
\end{eqnarray}
where $A_i$ are arbitrary constants. This generalized logotropic EoS is
interesting in its own right and, as we shall see, possesses interesting
properties. However, the above derivation suggests that the standard logotropic EoS (\ref{eq8}) plays a special role in the problem.

\section{Generalized Logotropic Cosmology}
\label{sec:cosm}

In this section, we consider a flat homogeneous and isotropic 
Friedmann-Lema\^itre-Robertson-Walker (FLRW) universe
filled with a perfect single fluid, having an energy density $\epsilon$,
rest-mass density $\rho_m$, and pressure $P$.
The expansion dynamics is governed by the Friedmann equations
\cite{weinberg2002}
\begin{eqnarray}
H^2  = \left (\frac{\dot a}{a}\right )^2= \frac{8 \pi G}{3 c^2} \epsilon, 
\label{eq9}
\end{eqnarray}
\begin{eqnarray}
\dot{H} + H^2 =  \frac{\ddot{a}}{a} = - \frac{4 \pi G}{3c^2} \left(\epsilon +
3P \right),
\label{eq9b}
\end{eqnarray}
where $a(t)$ is the scale factor and $H=\dot{a}/a$ is the Hubble
parameter. Combining Eqs. (\ref{eq9}) and (\ref{eq9b}), we obtain the energy
conservation equation
\begin{equation}
\label{fe0}
\frac{d\epsilon}{dt}+3H\left
(\epsilon+P\right )=0.
\end{equation}

For a relativistic fluid with an adiabatic evolution (or at $T=0$), the first
law of
thermodynamics reduces to
\begin{eqnarray}
d \epsilon =  \frac{P + \epsilon}{\rho_m} d \rho_m.
\label{eq10}
\end{eqnarray}
By integrating this equation for a given EoS, $P=P(\rho_m)$, a relation between
the energy density and the rest-mass
density can be obtained as \cite{chavanis2015}
\begin{eqnarray}
\epsilon  =  \rho_m c^2 + \rho_m \int^{\rho_m} \frac{P
\left(\rho'\right)}{\rho{'}^2} d \rho'=\rho_m c^2+u(\rho_m),
\label{eq11}
\end{eqnarray}
where $\rho_m c^2$ is the rest-mass energy density and $u$ is the internal
energy density.
For our generalized logotropic model with an EoS given by Eq. (\ref{eq1}), we
get
\begin{eqnarray}
\epsilon  =  \rho_m c^2 - \sum_{i=0}^N A_i I_i \left( \rho_m \right)
\label{eq12}
\end{eqnarray}
with the integral $I_i \left( \rho_m \right)$ given by
\begin{eqnarray}
I_i \left( \rho_m \right)  =  \rho_m\int_{\rho_m}^{+\infty}\ln^i\left
(\frac{\rho'_m}{\rho_P}\right )\, \frac{d\rho'_m}{{\rho'}^2}.
\end{eqnarray}
With the change of variables $t=\ln(\rho'_m/\rho_P)$, we obtain
\begin{eqnarray}
I_i \left( \rho_m \right)  = 
\frac{\rho_m}{\rho_P}\int_{\ln(\rho_m/\rho_P)}^{+\infty}t^i e^{-t}\, dt.
\end{eqnarray}
In terms of the  incomplete gamma function
\begin{eqnarray}
\Gamma(\alpha+1,x)=\int_x^{+\infty} t^{\alpha}e^{-t}\, dt
\label{gamma}
\end{eqnarray}
the integral $I_i \left( \rho_m \right)$ can be rewritten as
\begin{eqnarray}
I_i \left( \rho_m \right)  = \frac{\rho_m}{\rho_P}
~\Gamma \left\lbrack i+1, \ln \left( \frac{\rho_m}{\rho_P} \right)
\right\rbrack =  i! \sum_{k=0}^i \frac{1}{k!} \ln^k \left(
\frac{\rho_m}{\rho_P} \right).
\label{eq13}
\end{eqnarray}
To get the second equality, we have used the relation
\begin{eqnarray}
\Gamma \left(n+1, x \right)=n!\, e^{-x}\sum_{k=0}^n \frac{1}{k!} x^k, 
\label{eq14}
\end{eqnarray}
which can be obtained by computing $\partial\Gamma(n+1,x)/\partial x$ from Eqs.
(\ref{gamma}) and (\ref{eq14}).
For future purposes, we note the asymptotic behavior\footnote{We have assumed
that $n$ is an integer. However, it can be shown that this asymptotic
behavior remains valid when $n$ is a real number.}
\begin{eqnarray}
\Gamma \left(n+1, x \right)\sim x^n e^{-x}\qquad (x\rightarrow\infty).
\label{id}
\end{eqnarray}
The energy
density can then be written as
\begin{eqnarray}
\epsilon=\rho_m c^2-\sum_{i=0}^{N} A_i \frac{\rho_m}{\rho_P} \Gamma\left\lbrack
i+1, \ln\left
(\frac{\rho_m}{\rho_P}\right )\right\rbrack.
\label{asw1}
\end{eqnarray}
Equations (\ref{eq1}) and (\ref{asw1}) determine the EoS $P(\epsilon)$ in
parametric form.
By combining the energy conservation equation (\ref{fe0}) with the first law of
thermodynamics (\ref{eq10}) one finds that the rest-mass density
evolves as \cite{chavanis2015}
\begin{eqnarray}
\frac{d\rho_m}{dt}+3H\rho_m=0\qquad \Rightarrow \qquad \rho_m  = 
\frac{\rho_{m,0}}{a^3},
\label{eq15}
\end{eqnarray}
where $\rho_{m,0}= \rho_m \left(a_0=1 \right)$ is the present value of the
rest-mass density and $a_0=1$ is the present value of the scale factor.
Equation 
(\ref{eq15}) expresses the conservation of the rest-mass. On the other hand,
the internal energy is given by
\begin{eqnarray}
u  =  - \sum_{i=0}^N A_i I_i \left( \rho_m \right)= - \sum_{i=0}^N A_i I_i
\left(  \frac{\rho_{m,0}}{a^3}\right).
\label{eq12b}
\end{eqnarray}
As argued in \cite{chavanis2015}, the rest-mass energy density $\rho_m c^2$
plays the role of pressureless DM ($\epsilon_m=\rho_m c^2$) and the internal
energy $u$ plays the role of DE
($\epsilon_{de}=u$).\footnote{The notation $\rho_m$ stands
either for
pressureless DM density or for rest-mass density.} The energy density $\epsilon
= \epsilon_m + \epsilon_{de}$ of the generalized
logotropic model is therefore the sum of two terms where the first term
$\epsilon_m$ can be interpreted as the DM energy density given by
\begin{eqnarray}
\epsilon_m  =  \frac{\epsilon_{m,0}}{a^{3}}
\label{eq16}
\end{eqnarray}
and the second term $\epsilon_{de}$ can be interpreted as the DE
energy density given by
\begin{eqnarray}
\epsilon_{de}  =  - \frac{e^{-1-1/B}}{a^3} \sum_{i=0}^N A_i ~\Gamma \left(i+1,
-1 - \frac{1}{B} - 3 \ln a \right),
\label{eq17}
\end{eqnarray}
where, following \cite{chavanis2015}, we have defined the dimensionless
parameter $B$ through the relation
\begin{eqnarray}
\frac{\rho_P}{\rho_{m,0}}  =  e^{1+ 1/B}.
\label{eq18}
\end{eqnarray}
The Friedmann equation
(\ref{eq9}) can be written as
\begin{eqnarray}
H^2  =  H_0^2 \left( \frac{\epsilon_m}{\epsilon_0} +
\frac{\epsilon_{de}}{\epsilon_0} \right),
\label{eq19}
\end{eqnarray}
where $\epsilon_0 = {3 H_0^2 c^2}/{8 \pi G}$ is the critical energy density and
$H_0$ is the value of the Hubble parameter at the present time. 

The pressure and the energy density of the generalized logotropic model can be
expressed in terms of the scale factor as
\begin{eqnarray}
P  = \sum_{i=0}^N A_i \left(-1 - \frac{1}{B} - 3 \ln a \right)^i,
\label{eq20}
\end{eqnarray}
\begin{eqnarray}
\epsilon =  \frac{\Omega_{m,0} \epsilon_0}{a^3} -
\frac{e^{-1- 1/B}}{a^3} \sum_{i=0}^N A_i ~\Gamma \left(i+1,-1 - \frac{1}{B} - 3
\ln a \right).
\label{eq20b}
\end{eqnarray}
The present proportion  of DM and DE, denoted  $\Omega_{m,0}$ and
$\Omega_{de,0}$, are given by
\begin{eqnarray}
\Omega_{m,0} =  \frac{\epsilon_{m,0}}{\epsilon_0}, 
\label{eq21}
\end{eqnarray}
\begin{eqnarray}
\Omega_{de,0} =  \frac{\epsilon_{de,0}}{\epsilon_0} =
- \frac{e^{-1 - 1/B}}{\epsilon_0} \sum_{i=0}^N A_i ~\Gamma \left(i+1,-1 -
\frac{1}{B} \right) = 1- \Omega_{m,0}.
\label{eq21b}
\end{eqnarray}
For given values of $\Omega_{m,0}$ and $\epsilon_0$ (through $H_0$), which can
be obtained from the observations, Eq. (\ref{eq18}) determines $B$ and
Eq. (\ref{eq21b}) provides a constraint on the
coefficients $A_i$. Using $\rho_{P}={c^5}/({\hbar
G^2})=5.16\times 10^{99}\, {\rm g\, m^{-3}}$ and
$\rho_{m,0}=\Omega_{m,0}\epsilon_0/c^2$ with $\Omega_{m,0}=0.309$ and
$\epsilon_0/c^2=8.62\times 10^{-24}\, {\rm g\, m^{-3}}$, we get
$B=3.53\times 10^{-3}$. We have taken the values of $\Omega_{m,0}$ and
$\epsilon_0$ from the $\Lambda$CDM model but since $B$ is defined by a
logarithm, its value is very insensitive to the precise values of $\Omega_{m,0}$
and $\epsilon_0$. Therefore, the value $B=3.53\times 10^{-3}$ is very robust
\cite{chavanis2016}.

In the late Universe  ($a\rightarrow \infty$ and $\rho \rightarrow 0$), the DE
dominates and, using Eq. (\ref{id}), we get $\epsilon \sim -A_N(-3\ln a)^N$ and 
$P\sim A_N(-3\ln a)^N$, which implies that the EoS $P (\epsilon)$
behaves asymptotically as $P/\epsilon\rightarrow -1$. We note that in order to
have $\epsilon>0$ when $a\rightarrow +\infty$, the parameter $A_N$ must be
negative if $N$ is even and positive if $N$ is odd. Below we present some
interesting UDM models derived from the generalized logotropic model,
focusing  on the EoS $P (\epsilon)$, the energy density $\epsilon$
and the cosmological implications of these models. We also investigate the era
of DE dominance that drives the accelerated expansion of the Universe.

\section{Particular models}
\label{sec_part}

\subsection{The case $A_i= A_n ~\delta_{i,n}$}

This case focuses on just the n-th order term of the finite series leading to
the EoS
\begin{eqnarray}
P  =    A_n \ln^n \left(\frac{\rho_m}{\rho_{P}} \right).
\label{eq1b}
\end{eqnarray}
The energy density is given by 
\begin{eqnarray}
\epsilon=\rho_m c^2-A_n \frac{\rho_m}{\rho_P} \Gamma\left\lbrack
n+1, \ln\left
(\frac{\rho_m}{\rho_P}\right )\right\rbrack=\rho_m
c^2+u=\epsilon_m+\epsilon_{de},
\end{eqnarray}
where the first term is the rest-mass energy density (DM) and the second term
is the internal energy density (DE). The pressure and
total energy
density as a function of the scale factor reduce to
\begin{eqnarray}
P  =    A_n \left (-1-\frac{1}{B}-3\ln a\right )^n,
\label{eq1c}
\end{eqnarray}
\begin{eqnarray}
\epsilon  = \frac{\Omega_{m,0} \epsilon_0}{a^3} - A_n \frac{e^{-1 -1/B}}{a^3}
~\Gamma \left(n+1, - 1 - \frac{1}{B} - 3 \ln a \right).
\label{eq22}
\end{eqnarray}
At the present time (i.e. $a =1$), the above equation leads to
\begin{eqnarray}
A_n  =  - \frac{\left(1 - \Omega_{m,0} \right) \epsilon_0 e^{1+ 1/B}}{\Gamma
\left(n+1, -1 - \frac{1}{B} \right)},
\label{eq23}
\end{eqnarray}
which determines $A_n$ as a function of the measured values of $\Omega_{m,0}$
and $\epsilon_0$. The dimensionless constant $B$ is also determined by the
measured values of $\Omega_{m,0}$
and $\epsilon_0$ (see above).  As a
result, there is no free parameter in this model. Note that
\begin{eqnarray}
\frac{A_n}{\left(1 - \Omega_{m,0} \right) \epsilon_0}  =  - \frac{
e^{1+ 1/B}}{\Gamma
\left(n+1, -1 - \frac{1}{B} \right)}
\label{eq23b}
\end{eqnarray}
is a dimensionless constant depending only on $n$ and $B$.
Furthermore, $\rho_{\Lambda}\equiv \left(1 -
\Omega_{m,0} \right) \epsilon_0/c^2=5.96\times 10^{-24}\, {\rm g\, m^{-3}}$
is the present value of the DE
density (it is equal to the cosmological density $\rho_{\Lambda}=\Lambda/(8\pi
G)$ in the $\Lambda$CDM model). From Eq. (\ref{eq23b}), we find that $A_n<0$
for $n$ even
and $A_n>0$ for $n$ odd.  





On the other hand, from
Eq. (\ref{eq1b}), the
pressure $P$ vanishes when
$\rho_m=\rho_P$. If $n$ is even, the pressure $P$ is always negative. If $n$ is
odd, the pressure $P$ is positive for $\rho_m>\rho_P$ and negative for
$\rho_m<\rho_P$. In practice, we consider a regime where $\rho_m\ll\rho_P$
because the logotropic model does not describe the early inflation. In that
case, the pressure is always negative. The EoS $P(\epsilon)$ is
given in the reversed form $\epsilon(P)$ by
\begin{eqnarray}
\epsilon=e^{\mp \left |\frac{P}{A_n}\right |^{1/n}} \rho_P c^2-A_n e^{\mp \left
|\frac{P}{A_n}\right |^{1/n}} \Gamma\left (
n+1, \mp \left |\frac{P}{A_n}\right
|^{1/n}\right ),
\end{eqnarray}
where the upper sign corresponds to the most relevant case $\rho_m<\rho_P$ and
the lower sign corresponds to $\rho_m>\rho_P$. Substituting Eq. (\ref{eq23})
into Eqs. (\ref{eq1c}) and (\ref{eq22}), we obtain after simple manipulations
\begin{eqnarray}
P =  - \frac{\left(1 - \Omega_{m,0} \right) \epsilon_0}{\Gamma \left(n+1, -1
- \frac{1}{B} \right)} e^{1+1/B} ~\left(-1 - \frac{1}{B} - 3 \ln a \right)^n,
\label{eq25}
\end{eqnarray}
\begin{eqnarray}
\epsilon  =  \frac{\Omega_{m,0} \epsilon_0}{a^3} +
\frac{\left(1 - \Omega_{m,0} \right) \epsilon_0}{a^3}  ~\frac{\Gamma
\left(n+1,-1- \frac{1}{B} - 3 \ln a \right)}{\Gamma \left(n+1, -1 - \frac{1}{B}
\right)}.
\label{eq24}
\end{eqnarray}
In the early Universe ($a\rightarrow 0$, $\rho_m \rightarrow +\infty$) the DM
$\epsilon_m$ dominates and we have $\epsilon\sim
\Omega_{m,0}\epsilon_0/a^3$, $P\sim A_n(-3\ln a)^n$
such that $P/\epsilon\rightarrow 0$ whereas, in the late Universe
($a\rightarrow \infty$, $\rho_m \rightarrow 0$), the DE $\epsilon_{de}$
dominates and we have $\epsilon\sim -A_n(-3\ln a)^n$, $P\sim A_n(-3\ln a)^n$
such that $P/\epsilon\rightarrow -1$. Furthermore, in order to have $\epsilon>0$
for $a\rightarrow +\infty$, the parameter $A_n$ must be negative if $n$ is even
and positive if $n$ is odd. As we have seen, this is guaranteed by Eq.
(\ref{eq23}). Finally, it is worth mentioning that for $B\rightarrow 0$
(corresponding to a form of semiclassical limit $\rho_P\rightarrow +\infty$ or
$\hbar\rightarrow 0$),
the $\Lambda$CDM model is recovered \cite{chavanis2016}. Indeed, using Eq.
(\ref{id}), we
find that
Eqs. (\ref{eq25}) and (\ref{eq24}) reduce to
\begin{eqnarray}
P =  -\left(1 - \Omega_{m,0} \right) \epsilon_0,
\label{eq25b}
\end{eqnarray}
\begin{eqnarray}
\epsilon =  \frac{\Omega_{m,0}\epsilon_0}{a^3} +(1 -
\Omega_{m,0})\epsilon_0.
\label{eq24b}
\end{eqnarray}

\subsection{The case $A_i= A_0 ~\delta_{i,0}$ }

For $n=0$, we recover the $\Lambda$CDM
model (interpreted as a UDM model) corresponding to the EoS
\begin{eqnarray}
P  =  A_0.
\label{leq1bo}
\end{eqnarray}
The energy density is given by 
\begin{eqnarray}
\epsilon=\rho_m c^2-A_0,
\end{eqnarray}
where the first term is DM and the second term is DE.  The 
total energy
density as a function of the scale factor reads
\begin{eqnarray}
\epsilon  = \frac{\Omega_{m,0} \epsilon_0}{a^3} -A_0.
\label{leq22o}
\end{eqnarray}
At the present time (i.e. $a =1$), Eq. (\ref{leq22o}) leads to
\begin{eqnarray}
A_0  = - \left(1 - \Omega_{m,0}  \right) \epsilon_0.
\label{leq23g}
\end{eqnarray}
We note that $A_0<0$. Numerically,
$A_0/c^2=-\rho_{\Lambda}=-5.96\times 10^{-24}\, {\rm g\, m^{-3}}$,  where
$\rho_{\Lambda}$ is
the cosmological density. The pressure $P$ and   the
energy density $\epsilon$ are then given by Eqs. (\ref{eq25b}) and
(\ref{eq24b}).  The pressure $P=-\rho_{\Lambda}c^2$ is constant and negative. As
the universe expands, starting from $+\infty$, the energy density
$\epsilon$ decreases and tends to a constant value $\rho_{\Lambda}c^2$. The DE
density $\epsilon_{de}=\rho_{\Lambda}c^2$ is constant.

\subsection{The case $A_i= A_1 ~\delta_{i,1}$ }
For $n=1$, we recover the original
logotropic model ~\cite{chavanis2015} corresponding to the EoS
\begin{eqnarray}
P  =    A \ln \left(\frac{\rho_m}{\rho_{P}} \right),
\label{eq1bo}
\end{eqnarray}
where we have noted $A_1=A$. The energy density is given by
\begin{eqnarray}
\epsilon=\rho_m c^2-A\left\lbrack 1+\ln\left (\frac{\rho_m}{\rho_P}\right
)\right\rbrack,
\label{nrj}
\end{eqnarray}
where the first term is DM and the second term is DE. The EoS $P(\epsilon)$ is
given in the reversed form $\epsilon(P)$ by
\begin{eqnarray}
\epsilon=e^{P/A} \rho_P c^2-A-P.
\end{eqnarray}
The
pressure and
total energy
density as a function of the scale factor read
\begin{eqnarray}
P  =  -  A \left (1+\frac{1}{B}+3\ln a\right ),
\label{eq1co}
\end{eqnarray}
\begin{eqnarray}
\epsilon  = \frac{\Omega_{m,0} \epsilon_0}{a^3} + A \left (\frac{1}{B}+3\ln
a\right ).
\label{eq22o}
\end{eqnarray}
At the present time (i.e. $a =1$), Eq. (\ref{eq22o}) leads to
\begin{eqnarray}
A  =  B\left(1 - \Omega_{m,0}  \right) \epsilon_0.
\label{eq23g}
\end{eqnarray}
We note that $A>0$. Numerically,
$A/c^2=B\rho_{\Lambda}=3.53\times 10^{-3} \rho_{\Lambda}=2.10\times 10^{-26}\,
{\rm g\, m^{-3}}$. The pressure $P$ and the energy density $\epsilon$ become
\begin{eqnarray}
P =  -  \left(1 -
\Omega_{m,0}  \right) \epsilon_0 \left (B+1+3B\ln a\right ),
\label{eq1cob}
\end{eqnarray}
\begin{eqnarray}
\epsilon =  \frac{\Omega_{m,0}\epsilon_0}{a^3} +\left(1 -
\Omega_{m,0} \right) \epsilon_0\left(
1+ 3 B \ln a \right).
\label{eq27}
\end{eqnarray}
The pressure $P$ is positive when $\rho_m>\rho_P$ and negative
when $\rho_m<\rho_P$. It vanishes at $\rho_m=\rho_P$. As the universe expands,
the pressure decreases from
$+\infty$ to $-\infty$. Starting from $+\infty$, the energy density
$\epsilon$ first
decreases, reaches a minimum $\epsilon_{\rm min}=A\ln(\rho_P c^2/A)>0$ at
$\rho_m=A/c^2$, then increases to $+\infty$. The DE
density $\epsilon_{de}$ increases from
$-\infty$ to $+\infty$. The DE is negative when $\rho_m>\rho_P/e$ and positive
when  $\rho_m<\rho_P/e$  (its value at $\rho_m=\rho_P$ is
$\epsilon_{de}=-A<0$). In the logotropic
model, since the DE density corresponds to the internal energy density $u$ of
the LDF, it can very well be negative as long as the total energy density
$\epsilon$ is positive. Note, however, that in the regime of interest
$\rho_m\ll\rho_P$, the DE density $\epsilon_{de}$ is positive.

\subsection{The case $A_i= A_2 ~\delta_{i,2}$}
\label{sec_casd}

For $n=2$, we obtain the EoS
\begin{eqnarray}
P  =    A_2 \ln^2 \left(\frac{\rho_m}{\rho_{P}} \right).
\label{eq1bd}
\end{eqnarray}
The energy density is given by
\begin{eqnarray}
\epsilon=\rho_m c^2-2A_2\left\lbrack 1+\ln\left(\frac{\rho_m}{\rho_{P}}
\right)+\frac{1}{2} \ln^2 \left(\frac{\rho_m}{\rho_{P}} \right)\right\rbrack,
\end{eqnarray}
where the first term is DM and the second term is DE. The EoS $P(\epsilon)$ is
given in the reversed form $\epsilon(P)$ by
\begin{eqnarray}
\epsilon=e^{\mp \sqrt{\frac{P}{A_2}}} \rho_P c^2-2A_2 \left
(1\mp\sqrt{\frac{P}{A_2}}+\frac{1}{2}\frac{P}{A_2}\right ),
\end{eqnarray}
where the upper sign corresponds to the most relevant case $\rho_m<\rho_P$ and
the lower sign corresponds to $\rho_m>\rho_P$. The pressure and
total energy
density as a function of the scale factor read
\begin{eqnarray}
P  =    A_2 \left (1+\frac{1}{B}+3\ln a\right )^2,
\label{eq1cd}
\end{eqnarray}
\begin{equation}
\epsilon=\frac{\Omega_{m,0}\epsilon_0}{a^3}-2A_2\left\lbrack
-\frac{1}{B}-3\ln a+\frac{1}{2}\left (1+\frac{1}{B}+3\ln a\right
)^2\right\rbrack.
\label{e28d}
\end{equation}
At the present time (i.e. $a =1$), Eq. (\ref{e28d}) leads to
\begin{eqnarray}
A_2  =  - \frac{1- \Omega_{m,0}}{1 + \frac{1}{B^2}} \epsilon_0.
\label{eq28}
\end{eqnarray}
We note that $A_2<0$. Numerically,
$A_2/c^2=-B^2/(1+B^2)\rho_{\Lambda}=-1.25\times
10^{-5}\rho_{\Lambda}=7.43\times 10^{-29}\,
{\rm g\, m^{-3}}$. The pressure $P$ and  the energy
density $\epsilon$
become
\begin{eqnarray}
P =  - \frac{\left(1 - \Omega_{m,0} \right) \epsilon_0}{1 + B^2}
~\left(1 + B + 3 B \ln a \right)^2,
\label{eq29}
\end{eqnarray}
\begin{eqnarray}
\epsilon  =  \frac{\Omega_{m,0} \epsilon_0}{a^3} +
\left(1 - \Omega_{m,0} \right) \epsilon_0 ~
\frac{1 + B^2 + 6 B \ln a + 9 B^2 \ln^2 a}{1+ B^2}.
\label{eq29b}
\end{eqnarray}
The pressure $P$ is always
negative and vanishes at $\rho_m=\rho_P$. As the universe expands,
starting from $-\infty$, 
the pressure first increases, vanishes at $\rho=\rho_m$, then decreases  to
$-\infty$. Starting from $+\infty$, the energy density $\epsilon$ first
decreases, reaches a minimum  $\epsilon_{\rm min}=-A_2(\rho_m
c^2/2A_2-1)^2>0$ at
$\rho_m$ solution of $\rho_m c^2=2A_2[1+\ln(\rho_m/\rho_P)]$, then
increases to $+\infty$.  Starting from $+\infty$, the DE
density $\epsilon_{de}$ first decreases, reaches a minimum $\epsilon_{de}^{\rm
min}=-A_2>0$ at $\rho_m=\rho_P/e$, then increases to $+\infty$ (its value at
$\rho_m=\rho_P$ is $\epsilon_{de}=-2A_2>0$).

Cases with $n=1,2$ are the simplest logotropic models. Since $n$ is a free
parameter, cases with $n>2$ lead to a collection of generalized logotropic
models.

\subsection{The case $N=2$}
This generalized logotropic model is obtained by considering the first three
terms $A_0$, 
$A_1$ and $A_2$. This leads to the EoS
\begin{eqnarray}
P  =  A_0+  A_1 \ln \left(\frac{\rho_m}{\rho_{P}} \right)+A_2 \ln^2
\left(\frac{\rho_m}{\rho_{P}} \right).
\label{eq1bt}
\end{eqnarray}
The energy density is given by
\begin{eqnarray}
\epsilon=\rho_m c^2-A_0-A_1\left\lbrack 1+\ln\left (\frac{\rho_m}{\rho_P}\right
)\right\rbrack-2A_2\left\lbrack 1+\ln\left(\frac{\rho_m}{\rho_{P}}
\right)+\frac{1}{2} \ln^2 \left(\frac{\rho_m}{\rho_{P}} \right)\right\rbrack,
\label{sand}
\end{eqnarray}
where the first term is DM and the other terms are DE. The EoS $P(\epsilon)$
can be obtained  in the reversed form $\epsilon(P)$ by eliminating
$\rho_m$ between Eqs. (\ref{eq1bt}) and (\ref{sand}).\footnote{We can  note
that Eq. (\ref{eq1bt}) is a second degree equation in $\ln
({\rho_m}/{\rho_{P}})$.}
The pressure and the energy density evolve
with the scale factor as
\begin{eqnarray}
P  = A_0+A_1\left (-1-\frac{1}{B}-3\ln a\right )+A_2 \left (-1-\frac{1}{B}-3\ln
a\right )^2,
\label{e30}
\end{eqnarray}
\begin{eqnarray}
\epsilon  =  \frac{\Omega_{m,0}\epsilon_0}{a^3}-A_0+A_1\left
(\frac{1}{B}+3\ln
a\right )-2A_2\left\lbrack -\frac{1}{B}-3\ln a+\frac{1}{2}\left
(1+\frac{1}{B}+3\ln a\right )^2\right\rbrack.
\label{e31}
\end{eqnarray}
At the present time (i.e. $a=1$), the above equation gives
\begin{eqnarray}
(1-\Omega_{m,0}) ~\epsilon_0  =-A_0+ \frac{A_1}{B}-A_2\left
(1+\frac{1}{B^2}\right ),
\label{e32}
\end{eqnarray}
which constrains one of the three free parameters. As a result, the model has
only two free parameters. As expected, in the late Universe ($a\rightarrow
\infty$,
$\rho \rightarrow 0$), the DE density $\epsilon_{de}$
dominates and we
have $P/\epsilon\rightarrow -1$. It is worth mentioning that in order to have
$\epsilon>0$, $A_2$ must be negative.
\section{Evolution of the generalized logotropic model}
\label{sec_evo}

\begin{figure}[htp]
\centering
  \includegraphics[width=.6\textwidth]{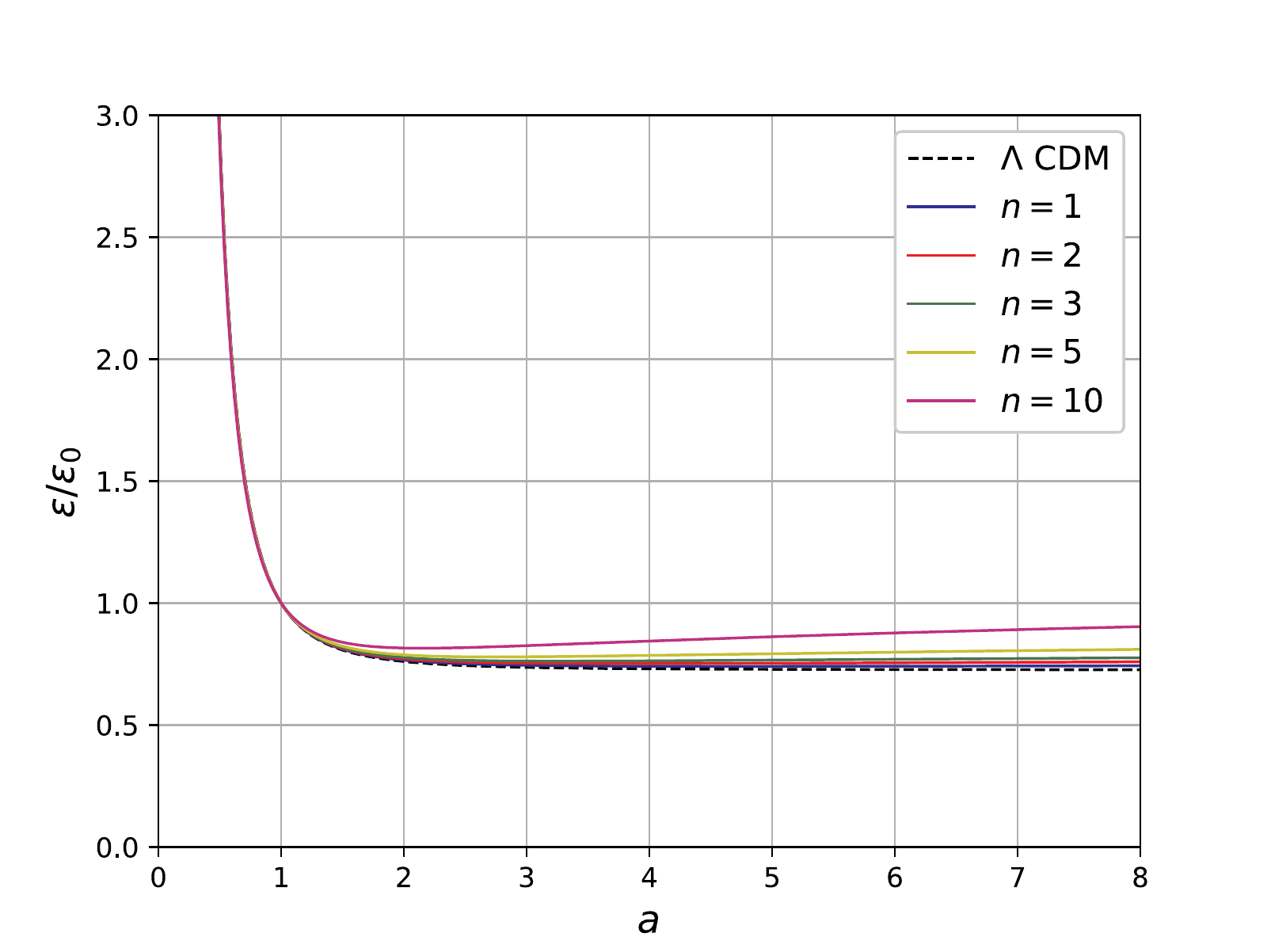} \caption{Energy density
$\epsilon/\epsilon_0$ as a function of the scale factor $a$ for
$\Omega_{m,0}=0.309$, $B=3.53 \times 10^{-3}$ and $n=0,1,2,3,5,10$}
 \label{fig1}
   \end{figure}

In the following we consider models with $A_i= A_n ~\delta_{i,n}$. The evolution
of the energy density as a function of the scale factor is plotted
in Fig. \ref{fig1},
where we have used the best-fit values of the parameters $\Omega_{m,0}
=0.309$ 
and
$B=3.53 \times 10^{-3}$  obtained from the $\Lambda$CDM model (they
are consistent with the cosmological analysis of the original
logotropic model \cite{chavanis2017, mamon2020}).  It shows that the Universe
starts at $a=0$ with an infinite energy
density.\footnote{We recall that the logotropic model is not aimed at describing
the
early inflation.} The
energy density $\epsilon$ first decreases with the
increase of the scale factor $a$, reaches a minimum, then 
increases with the scale factor characterizing a phantom Universe
\cite{caldwell,ckw}.

In the generalized logotropic model, the Friedmann equation (\ref{eq19}) takes
the form
\begin{eqnarray}
H  =  \frac{\dot{a}}{a} = H_0 \sqrt{\frac{\Omega_{m,0}}{a^3} +
\frac{1 - \Omega_{m,0}}{a^3}  ~\frac{\Gamma \left(n+1,-1-
\frac{1}{B} - 3 \ln a \right)}{\Gamma \left(n+1, -1 - \frac{1}{B} \right)} }.
\label{eq33}
\end{eqnarray}
 
\begin{figure}
\includegraphics[width=.6\textwidth]{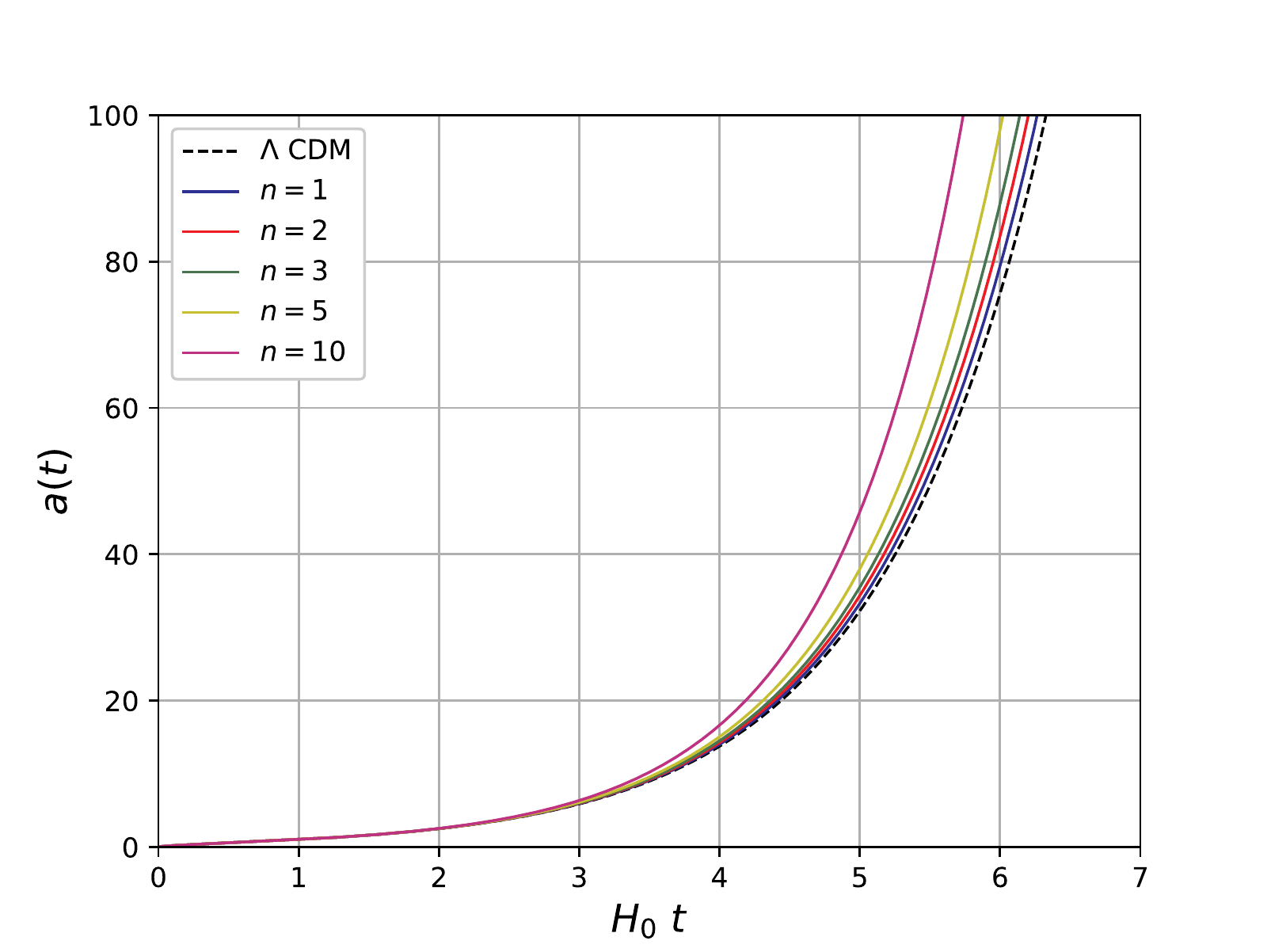} \caption{Scale factor $a$ as a
function of the cosmic time $t$ for $\Omega_{m,0}=0.309$, $B=3.53 \times
10^{-3}$ and
$n=0,1,2,3,5,10$}
\label{fig3}
\end{figure}

The evolution of
the scale factor as a function of the time is given by
\begin{eqnarray}
H_0 t  =  \int_0^a \frac{d a'}{a' ~\sqrt{\frac{\Omega_{m,0}}{a'^3} +
\frac{1 - \Omega_{m,0}}{a'^3}  ~\frac{\Gamma \left(n+1,-1-
\frac{1}{B} - 3 \ln a' \right)}{\Gamma \left(n+1, -1 - \frac{1}{B} \right)} } }.
\label{eq34}
\end{eqnarray}
The $\Lambda$CDM model is recovered from Eq. (\ref{eq34}) for $n=0$ or $B=0$. In
that
case, Eq. (\ref{eq34}) can be integrated analytically yielding
\begin{eqnarray}
a=\left (\frac{\Omega_{m,0}}{1-\Omega_{m,0}}\right )^{1/3}\sinh^{2/3}\left
(\frac{3}{2}\sqrt{1-\Omega_{m,0}}H_0 t\right ),
\end{eqnarray}
\begin{eqnarray}
\frac{\epsilon}{\epsilon_0}=\frac{1-\Omega_{m,0}}{\tanh^2\left
(\frac{3}{2}\sqrt{1-\Omega_{m,0}}H_0 t\right )},
\end{eqnarray}
whereas for $B \neq 0$ it
can only be integrated numerically. Figure \ref{fig3} shows the behavior of the
scale
factor $a$ as a function of $H_0 t$ for $n=0,1,2,3,5,10$. 
The age of the universe is given by
\begin{eqnarray}
 t_{\rm age}  = \frac{1}{H_0} \int_0^1 \frac{d a'}{a'
~\sqrt{\frac{\Omega_{m,0}}{a'^3} +
\frac{1 - \Omega_{m,0}}{a'^3}  ~\frac{\Gamma \left(n+1,-1-
\frac{1}{B} - 3 \ln a' \right)}{\Gamma \left(n+1, -1 - \frac{1}{B} \right)} } }
\label{eq34age}
\end{eqnarray}
with $H_0=2.195\times 10^{-18}\, {\rm s}^{-1}$, i.e.,  $H_0^{-1}=14.4\, {\rm
Gyrs}$.
For the $\Lambda$CDM model ($n=0$) we recover the well-known result $t_{\rm
age}=0.956\, H_0^{-1}=13.8\, {\rm Gyrs}$. A difference larger than
$0.1\,\%$ (the
typical error bar on the age of the universe) occurs in models with $n>3$
so these models should be rejected.  For example, we get $t_{\rm
age}=0.966\, H_0^{-1}=13.9\, {\rm Gyrs}$ for $n=20$, $t_{\rm
age}=0.998\, H_0^{-1}=14.4\, {\rm Gyrs}$ for $n=100$, and  $t_{\rm
age}=1.12\, H_0^{-1}=16.1\, {\rm Gyrs}$ for $n=1000$.

The asymptotic behavior of the generalized logotropic model can be obtained
analytically.
For $a\rightarrow 0$, we have a matter dominated Universe which corresponds to
Einstein-deSitter solution given by
\begin{eqnarray}
a \sim  \left(\frac{3}{2} \sqrt{\Omega_{m,0}} H_0 t \right)^{2/3},
\label{eq35}
\end{eqnarray}
\begin{eqnarray}
\frac{\epsilon}{\epsilon_0} \sim \frac{4}{9 H_0^2 t^2}.
\label{eq36}
\end{eqnarray}
For $a\rightarrow +\infty$, the energy density behaves as
\begin{eqnarray}
\epsilon\sim 3^n |A_n|(\ln a)^n.
\label{e37}
\end{eqnarray}
In this asymptotic regime, the Friedmann equation can be written as
\begin{eqnarray}
\frac{\dot a}{a}=\sqrt{\frac{8\pi G}{3c^2} \epsilon} \sim K_n ~(\ln a)^{n/2},
\label{e38}
\end{eqnarray}
where we define the constant $K_n$ as
\begin{eqnarray}
K_n   =  3^{n/2} H_0 \sqrt{\frac{\left(1 - \Omega_{m,0} \right)
e^{1+1/B}}{\Gamma \left(n+1, -1 - \frac{1}{B} \right)}}.
\label{e39}
\end{eqnarray}
The above equation can be integrated into
\begin{eqnarray}
K_n ~t \sim \int^a \frac{da'}{a' (\ln a')^{n/2}} \sim\int^{\ln
a}\frac{ds}{s^{n/2}},
\label{e40}
\end{eqnarray}
where we have made the change of variables $s=\ln a'$ to get the second
equality. It is interesting to note that the asymptotic evolution of the scale
factor as a
function of time depends mainly on the value of the parameter $n$. We can
distinguish three relevant types of evolution which correspond to $n<2, n=2$ and
$n>2$.

(i) For $n<2$: this solution, which includes the
original logotropic model
\cite{chavanis2015,chavanis2016,chavanis2017}, describes a super de Sitter
evolution of the form
\begin{eqnarray}
a \propto  {\rm exp}\left\lbrace {\left\lbrack\frac{1}{2}(2-n) ~K_n~
t\right\rbrack^{\frac{2}{2-n}}}\right\rbrace,\qquad \epsilon\propto t^n,
\label{e41}
\end{eqnarray}
where the scale factor grows super exponentially rapidly with cosmic time
causing an
algebraic divergence of the energy density. For $n=1$ we
recover the original logotropic
model where $a\sim e^{t^2}$ and $\epsilon\sim
t$ \cite{chavanis2015,chavanis2016,chavanis2017}. Since the scale
factor and the density increase indefinitely with time, this is called
Little Rip \cite{littlerip}. For $n=0$ we
recover the $\Lambda$CDM model presenting an exponential (de Sitter) expansion
$a\sim e^t$ and a constant energy density $\epsilon\sim 1$.

(ii) For $n=2$: we get  a double exponential
evolution of the form
\begin{eqnarray}
a \propto {\rm exp}{\lbrack{{\rm exp} (K_2 ~t) \rbrack}},\qquad \epsilon\propto
{\rm exp}(2K_2 t),
\label{e42}
\end{eqnarray}
where the scale factor grows hyper exponentially rapidly with cosmic time
causing an
exponential divergence of the energy density (Little Rip).

(iii) For $n>2$: this case represents a situation
in which the Universe ends up
with a finite-time future singularity. One finds
\begin{eqnarray}
a  \propto  {\rm exp}\left\lbrace {\left\lbrack\frac{1}{2}(n-2) ~K_n
~(t_s-t)\right\rbrack^{-\frac{2}{n-2}}}\right\rbrace,\qquad
\epsilon\propto \left\lbrack\frac{1}{2}(n-2) ~K_n
~(t_s-t)\right\rbrack^{-\frac{2n}{n-2}}.
\label{e43}
\end{eqnarray}
The singularity at $t=t_s$ corresponds to a Big Rip \cite{caldwellprl}
characterized by the
divergence of the scale factor and energy density in finite time. 
The big rip time is given by
\begin{eqnarray}
t_{\rm BR}  = \frac{1}{H_0} \int_0^{+\infty} \frac{d a'}{a'
~\sqrt{\frac{\Omega_{m,0}}{a'^3} +
\frac{1 - \Omega_{m,0}}{a'^3}  ~\frac{\Gamma \left(n+1,-1-
\frac{1}{B} - 3 \ln a' \right)}{\Gamma \left(n+1, -1 - \frac{1}{B} \right)} } }.
\label{eq34br}
\end{eqnarray}
For measured values of $\Omega_{m,0}$ and $\epsilon_0$ (hence $H_0$), this is
just a function of $n$. There is no other free
(undetermined) parameter in our model. We get $t_{\rm BR}=3240\, {\rm
Gyrs}$ for $n=3$, $t_{\rm BR}=1650\, {\rm
Gyrs}$ for $n=4$, $t_{\rm BR}=422\, {\rm
Gyrs}$ for $n=10$,  $t_{\rm BR}=47.1\, {\rm
Gyrs}$ for $n=100$, and $t_{\rm BR}=19.2\, {\rm
Gyrs}$ for $n=1000$. However, we recall that the models
with $n>3$ are excluded from the observations. 



\begin{figure}
\centering
  \includegraphics[width=.6\textwidth]{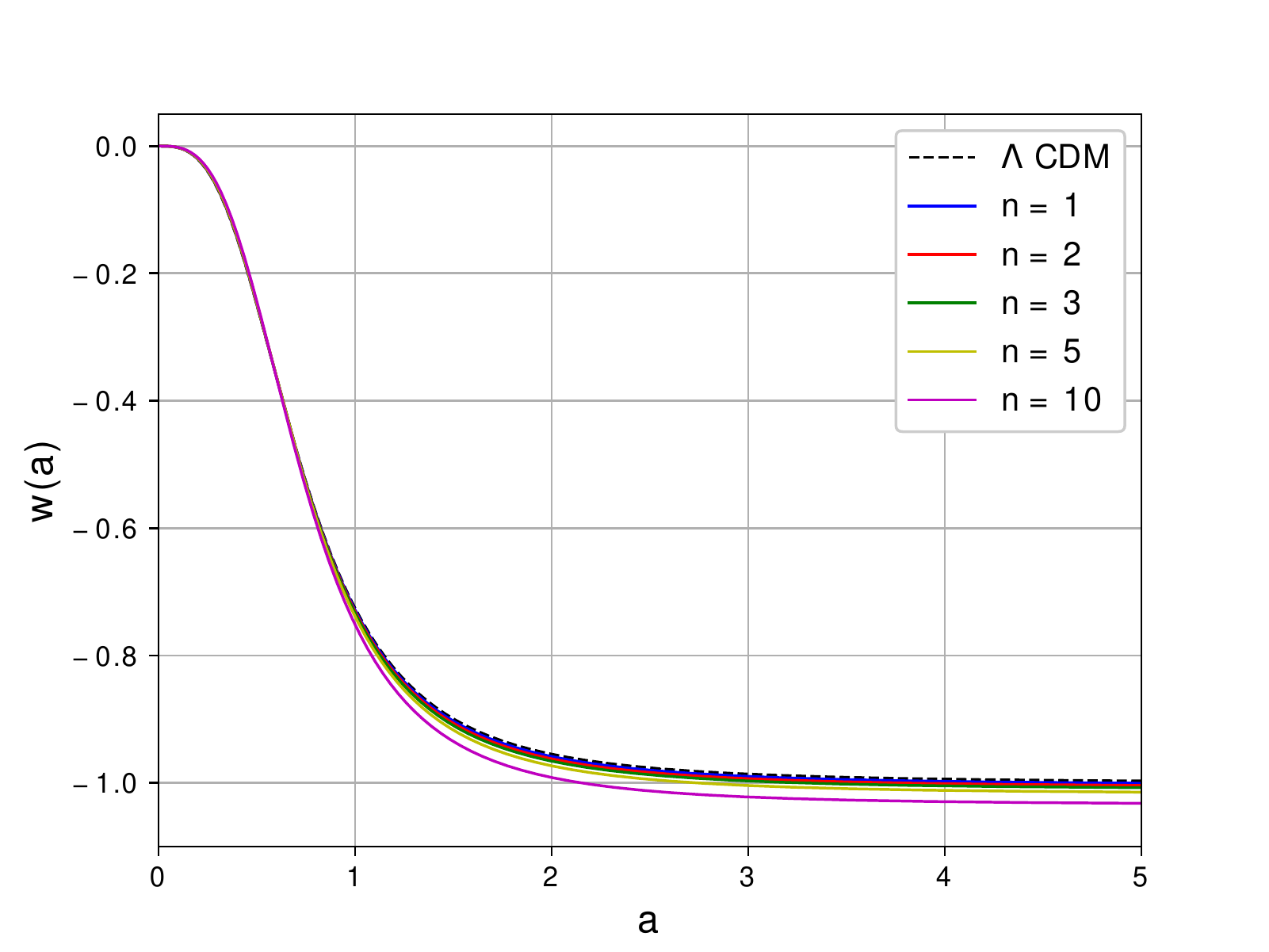} \caption{EoS parameter $w
(a)$ as a function the scale factor $a$ for $\Omega_{m,0}=0.309$,
$B=3.53 \times
10^{-3}$ and $n=0,1,2,3,5,10$}
\label{fig4}
\end{figure}

Using Eqs. (\ref{eq25}) and (\ref{eq24}), the EoS
parameter $w=P/\epsilon$  can be expressed in terms of the scale factor as
\begin{eqnarray}
w (a)  =  \frac{-\left(1 - \Omega_{m,0} \right) e^{1+1/B} a^3
\left
(-1-\frac{1}{B}-3\ln a\right )^n}
{\Omega_{m,0} \Gamma \left(n+1,-1- \frac{1}{B} \right) + \left(1 - \Omega_{m,0}
\right) \Gamma \left(n+1,-1- \frac{1}{B} - 3 \ln a \right)}.
\label{eq44}
\end{eqnarray}
For $a=1$, we get
\begin{eqnarray}
w_0  = - \frac{\left(1 - \Omega_{m,0} \right) e^{1+1/B}  \left
(-1-\frac{1}{B}\right )^n}{ \Gamma \left (n+1,-1- \frac{1}{B} \right)}.
\label{eq44bw}
\end{eqnarray}
For the $\Lambda$CDM ($n=0$) we recover the well-known value $w_0=-\left(1 -
\Omega_{m,0} \right)=-0.691$. We get
$w_0=-0.693$ for $n=1$, $w_0=-0.696$ for $n=2$,
$w_0=-0.698$ for $n=3$, $w_0=-0.701$ for $n=4$, $w_0=-0.715$ for
$n=10$, $w_0=-0.935$ for $n=100$, and $w_0=-3.12$ for $n=1000$. For $n>3$
we are out of the error bars (typically $1\,\%$ for the value of $w_0$) so
these models should be rejected. The
behavior of the EoS parameter $w (a)$ as function of
the scale factor $a$ is
plotted in Fig. \ref{fig4}.
It shows that the GLDF behaves asymptotically as a superposition
of two non-interacting fluids (i.e. DM and DE) with the total
pressure and total
energy given by the sum 
\begin{eqnarray}
P  =  P_m + P_{de},\qquad 
\epsilon  = \epsilon_m + \epsilon_{de}.
\label{eq45}
\end{eqnarray}
Their EoS parameters are
\begin{eqnarray}
w_m  =  \frac{P_m}{\epsilon_m} = 0,\qquad 
w_{de}  =  \frac{P_{de}}{\epsilon_{de}}= - \frac{e^{1+1/B} a^3 \left
(-1-\frac{1}{B}-3\ln a\right )^n}
{\Gamma \left(n+1,-1- \frac{1}{B} - 3 \ln a \right)}.
\label{eq46}
\end{eqnarray}

\begin{figure}
\centering
  \includegraphics[width=.6\textwidth]{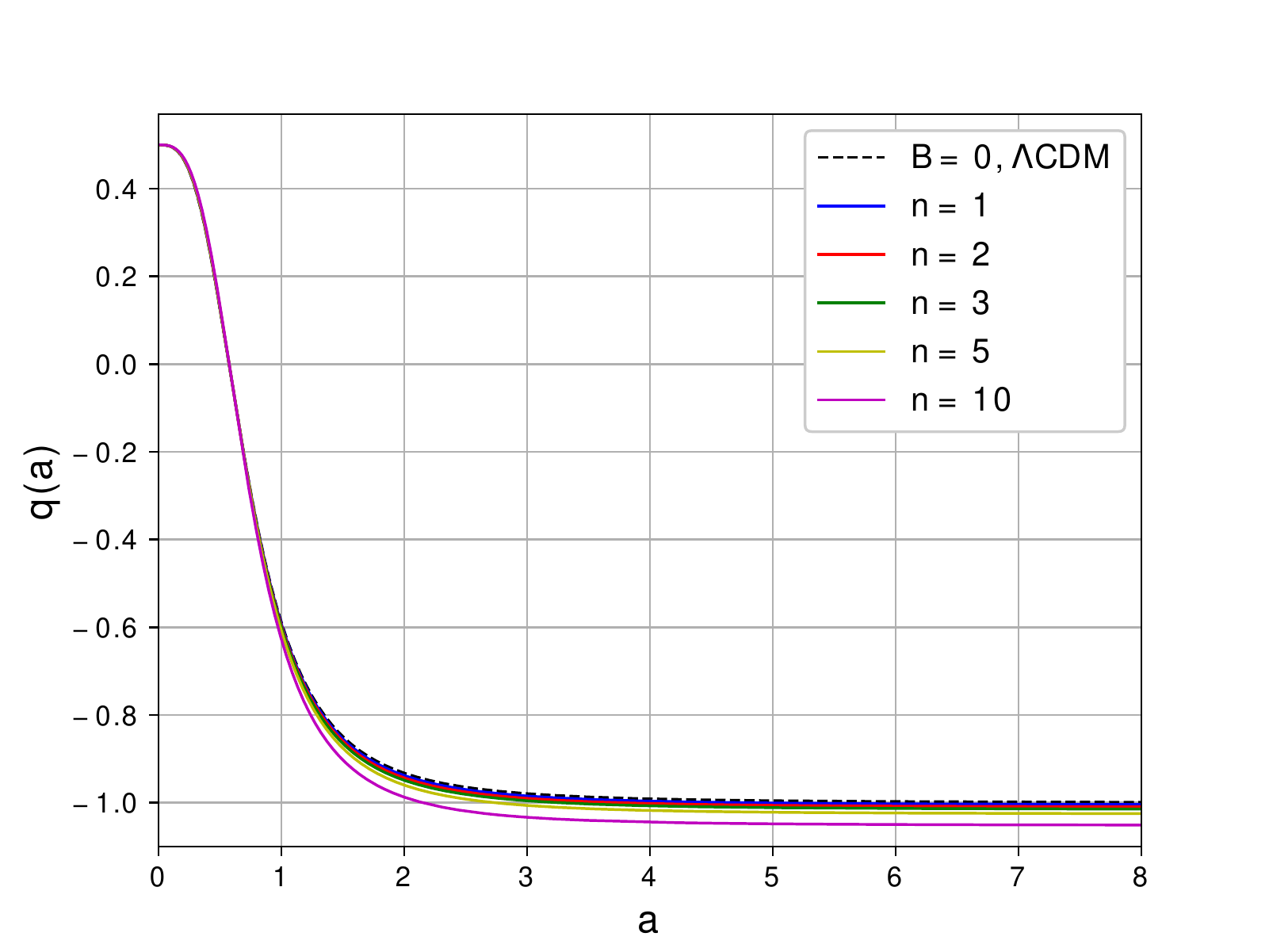} \caption{Acceleration $q
(a)$ as a function of the scale factor $a$ for $\Omega_{m,0}=0.309$,
$B=3.53 \times
10^{-3}$ and $n=0,1,2,3,5,10$}
\label{fig5}
    \end{figure}

In a flat Universe, the deceleration parameter $q=-{\ddot{a} a}/{\dot{a}^2}
= -{\dot{H}}/{H^2} - 1$ is related to the EoS parameter by
\begin{eqnarray}
q =\frac{1+3w}{2}.
\label{eq47}
\end{eqnarray}
The Universe is undergoing an accelerating expansion if $q < 0$ (i.e. $w<-1/3$).
In
Fig. \ref{fig5}, we show the behavior of $q (a)$ as a function of the scale
factor for
different values of $n$. 

From Eq. (\ref{eq24}) the transition scale factor
$a_t$, corresponding to $\epsilon_m=\epsilon_{de}$, is
obtained by solving the transcendental equation
\begin{eqnarray}
\Gamma \left(n+1,-1- \frac{1}{B} - 3 \ln a_t \right)  =  \frac{\Omega_{m,0}}{1
- \Omega_{m,0}}
~\Gamma \left(n+1,-1- \frac{1}{B} \right).
\label{eq48}
\end{eqnarray}

\begin{figure}
\centering
  \includegraphics[width=.6\textwidth]{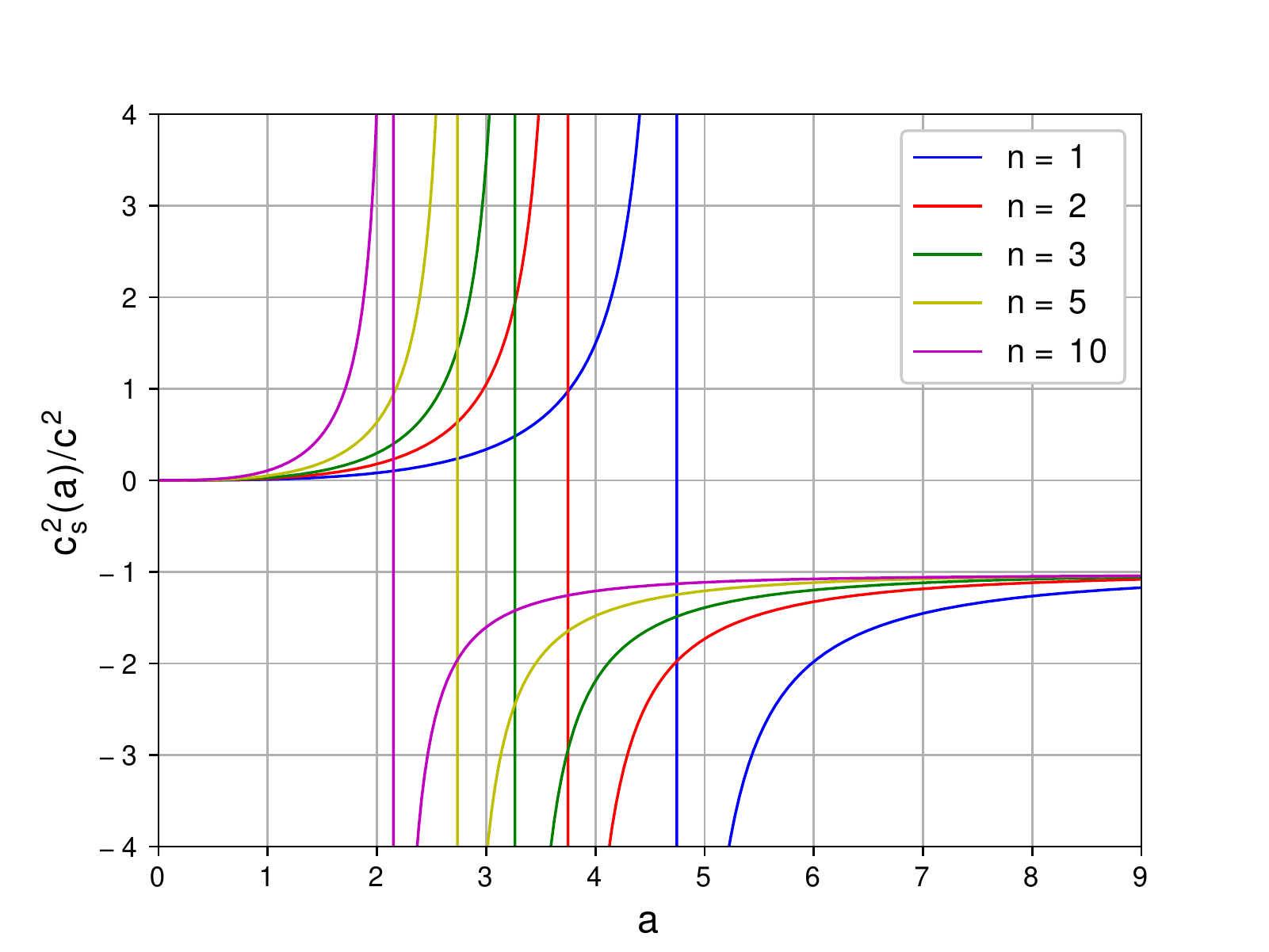} \caption{Squared speed of
sound $c_s^2 (a)/c^2$ as a function of the  scale factor $a$ for
$\Omega_{m,0}=0.309$, $B=3.53
\times 10^{-3}$ and $n=0,1,2,3,5,10$}
\label{fig6}
    \end{figure}

Another parameter to study is the squared  speed of sound $c_s^2$ which is a key
ingredient to investigate the stability of any model. In particular, the sign of
$c_s^2$ plays a crucial role for determining classical stability. It is defined
by
\begin{eqnarray}
c_s^2  =  \frac{dP}{d \epsilon} ~c^2 = \frac{dP/d \rho_m}{d \epsilon/d \rho_m}
~c^2.
\label{eq49}
\end{eqnarray}
Differentiating Eqs. (\ref{eq1}) and (\ref{eq12}) with $A_i= A_n
~\delta_{i,n}$, we get
\begin{eqnarray}
\frac{c_s^2}{c^2}  =  \frac{n\ln^{n-1} \left(\frac{\rho_m}{\rho_P}
\right)}{\frac{\rho_m c^2}{A_n} - n \frac{\rho_m}{\rho_P} \Gamma \left\lbrack n,
\ln
\left(\frac{\rho_m}{\rho_P} \right) \right\rbrack}.
\label{eq50}
\end{eqnarray}
In Fig. \ref{fig6} we plot the squared speed of sound $c_s^2/c^2$ as a function
of the
scale factor for different values of $n$. The causality and classical stability
conditions are satisfied if the speed of sound varies  in the range $0 \le
c_s^2/c^2 \le
1$.
One sees from Fig. \ref{fig6} that, in fact, there are regions where the
causality and
classical stability conditions are satisfied but the extent of these
regions decreases as $n$ increases.\footnote{The speed of sound is real in the
normal regime $d\epsilon/d\rho_m>0$ and imaginary in the phantom
regime $d\epsilon/d\rho_m<0$. It becomes infinite (before becoming imaginary)
when we enter into the phantom regime, i.e., when $d\epsilon/d\rho_m=0$.} 
\section{Conclusions}
\label{sec_con}

In this work, we have proposed a new class of cosmological unified dark sector
models ``{\em Generalized Logotropic Models}". These models are a generalization
of the logotropic model \cite{chavanis2015} by considering the
pressure $P$ as a sum of
higher logarithmic terms of the rest-mass density $\rho_m$.
The pressure can naturally take negative values in these cosmological
models. In
this scenario, the Universe is filled with a single fluid without the need of a
cosmological constant. Our generalized model depends on a set of free parameters
$A_i$ with $i=1,2, \cdots ,N$. In particular, we have considered a special
class of
generalized logotropic models where $A_i = A_n \delta_{in}$, which depends on
two free parameters $A_n$ and $n$. The usual logotropic model 
\cite{chavanis2015} 
corresponds to $n=1$
and $A_1$. We have also presented the model with $N=2$, which contains the first
two terms $A_1$ and $A_2$ of the series. We have highlighted the most relevant
properties of
these generalized logotropic models. To fix bounds on the free parameters of our
models, we employed the best fit of the parameters $\Omega_{m,0}$ and $B$
obtained from the cosmological analysis carried on the original logotropic model
(i.e. $n=1$) \cite{chavanis2017}. After fixing the free parameters, we
investigated the cosmological behavior of the generalized logotropic models by
focusing on the evolution of the DE density, scale factor, EoS
parameter, decceleration parameter and squared speed of sound. We
showed the
asymptotic behavior of these models and noticed three distinct ways of evolution
depending on the value of $n$.
In all the analyzed cases, we established that generalized logotropic models
lead to realistic cosmological models in which the dark sector is represented by
a unique fluid.
The deviation of the generalized models from the standard $\Lambda$CDM model
depends on the value of the parameter $n$. At later times, higher values of $n$
lead to higher deviations from $\Lambda$CDM. This implies that the generalized
logotropic models can be used as realistic  background cosmological models to
describe our Universe with a free parameter $n$. We estimated that only models
with $n\le 3$ are consistent with the observations. To find out the most
suitable
values of $n$, it will be necessary to perform a fitting procedure by using, for
instance, the Monte Carlo method and  a detailed comparison with cosmological
data like SN, BAO, and CMB surveys. We expect to perform this analysis in future
works. 


A further interesting generalization of the  logotropic model is to consider a
single fluid described by the EoS 
\begin{eqnarray}
P = A \ln^{\alpha} \left(\frac{\rho_m}{\rho_P} \right),
\end{eqnarray}
where $A$ is a real number given by
\begin{eqnarray}
A  =  - \frac{\left(1 - \Omega_{m,0} \right) \epsilon_0 e^{1+ 1/B}}{\Gamma
\left(\alpha +1, -1 - \frac{1}{B} \right)},
\label{eq52}
\end{eqnarray}
and $\alpha$ is a  real positive  parameter which can be constrained by the
cosmological data. There are two ways to recover the  $\Lambda$CDM, either by
taking $\alpha = 0$ or $B=0$ which makes the model rich and particularly more
appealing in describing DE especially for $0< \alpha < 1$. Such a model
interpolates between the $\Lambda$CDM for values of $\alpha$ close to zero and
to the
usual logotropic model for values of $\alpha$ close to one. 

In summary, single fluids with generalized logotropic EoS may
have interesting cosmological features and, thus, they represent a good
candidate to describe the DE sector. The investigation and
analysis of these models will be carried out in detail in future works.\\

Finally, we would like to mention that the logotropic model and the generalization presented in this work are not free of intrinsic problems, as all the cosmological models known in the literature. In fact, the speed of sound in logotropic models has the unpleasant property of increasing with the scale factor,  leading, like for the Chaplygin gas model, to oscillations in the mass power-spectrum that are not detected in observations at the cosmological level \cite{sandvik2004end}. However, there are several possibilities to solve this problem by considering additional effects such as non-linear and non-adiabatic perturbations, among others 
(see the discussion in \cite{chavanis2022new}).  
In any case, UDM models constitute a subject of intensive research as possible alternative scenarios to the popular and generally accepted $\Lambda$CDM model.  Our paper provides  a class of models exhibiting a transition between a normal behavior and a phantom behavior governed by a single equation of state. In addition, depending on the value of the parameter $n$, we can have different types of late evolution: no singularity, little rip, big rip... It is very interesting to note that all these models are consistent with the $\Lambda$CDM model up to the present time but will differ in the future. A virtue of our model is to show that it is very difficult to predict the future evolution of the universe based on present observations.


\section*{Acknowledgements}

We would like to thank Prof. O. Luongo for helpful comments.
HBB gratefully acknowledges the financial support from University of Sharjah
(grant number V.C.R.G./R.438/2020). The work of HQ was partially supported  by
UNAM-DGAPA-PAPIIT, Grant No. 114520, and Conacyt-Mexico, Grant No. A1-S-31269.

\appendix

\section{Asymptotic equation of state}
\label{sec_ae}

In this Appendix, we establish the asymptotic EoS $P(\epsilon)$ of the GLDF and
make the connection with the MCG model
\cite{benaoum2002,cosmopoly1,cosmopoly2,cosmopoly3}.

For $a\rightarrow +\infty$, using Eqs. (\ref{id}) and (\ref{eq20b}), we find
that the energy density evolves with the scale factor as
\begin{equation}
\epsilon\sim |A_N|\left (1+\frac{1}{B}+3\ln a\right )^N.
\label{ae1}
\end{equation}
Let us determine the corresponding asymptotic EoS from the energy
conservation equation (\ref{fe0}) which can be rewritten as
\begin{equation}
\frac{d\epsilon}{da}+\frac{3}{a}(\epsilon+P)=0.
\label{ae3}
\end{equation}
From Eqs. (\ref{ae1}) and (\ref{ae3}), we get
\begin{eqnarray}
P&=&-\epsilon-\frac{a}{3}\frac{d\epsilon}{da}\nonumber\\
&=&-\epsilon- N |A_N|\left (1+\frac{1}{B}+3\ln a\right )^{N-1}\nonumber\\
&=&-\epsilon- N |A_N|\left (\frac{\epsilon}{|A_N|}\right )^{(N-1)/N}\nonumber\\
&=&-\epsilon\left \lbrack 1+ N \left (\frac{|A_N|}{\epsilon}\right
)^{1/N}\right\rbrack.
\label{ae3b}
\end{eqnarray}
Therefore, the asymptotic EoS of the GLDF reads
\begin{eqnarray}
P=-\epsilon-N|A_N|^{1/N}\epsilon^{1-1/N}.
\label{ae3c}
\end{eqnarray}
This is a
particular case of the
generalized polytropic EoS (or MCG
model) \cite{benaoum2002,cosmopoly1,cosmopoly2,cosmopoly3}
\begin{eqnarray}
P=\alpha\epsilon+K\left (\frac{\epsilon}{c^2}\right )^{\gamma}
\label{ae3d}
\end{eqnarray}
corresponding to $\alpha=-1$, $K=-N(|A_N|c^2)^{1/N}$ and $\gamma=1-1/N$.
Therefore, the GLDF is asymptotically equivalent to the MCG with
$\alpha=-1$. Since $w=P/\epsilon<-1$, the EoS (\ref{ae3c}) leads to a phantom
behavior in agreement with the results of Sec. \ref{sec_evo}. For $N=1$, the
EoS (\ref{ae3c}) reduces to
\begin{eqnarray}
P=-\epsilon-A.
\label{ae10}
\end{eqnarray}
For $N=2$, it reduces to
\begin{eqnarray}
P=-\epsilon-2\sqrt{|A_2|}\epsilon^{1/2}.
\label{ae12}
\end{eqnarray}

Let us check that we recover the asymptotic EoS (\ref{ae3c})  directly from
the generalized logotropic EoS defined by Eqs. (\ref{eq20}) and (\ref{eq20b}).
For $a\rightarrow +\infty$, we
have
\begin{equation}
P\simeq A_N x^N+A_{N-1}x^{N-1}+...,
\label{ae4}
\end{equation}
\begin{equation}
\epsilon\simeq -A_N x^N-A_{N}Nx^{N-1}-A_{N-1}x^{N-1}+...,
\label{ae5}
\end{equation}
where
\begin{equation}
x\equiv -1-\frac{1}{B}-3\ln a\rightarrow -\infty.
\label{ae6}
\end{equation}
We stress that it is necessary to account for the first order correction to the
leading
term $x^N$ in Eqs. (\ref{ae4}) and (\ref{ae5}). From
\begin{equation}
P\simeq A_N x^N\left (1+\frac{A_{N-1}}{A_N}\frac{1}{x}\right ),
\label{ae7}
\end{equation}
\begin{equation}
\epsilon\simeq -A_N x^N\left
(1+N\frac{1}{x}+\frac{A_{N-1}}{A_N}\frac{1}{x}\right ),
\label{ae8}
\end{equation}
we get
\begin{eqnarray}
\frac{P}{\epsilon}&\simeq& -\left (1+\frac{A_{N-1}}{A_N}\frac{1}{x}\right )\left
(1-N\frac{1}{x}-\frac{A_{N-1}}{A_N}\frac{1}{x}\right )\nonumber\\
&\simeq& -\left (1-\frac{N}{x}\right )\nonumber\\
&=&-\left \lbrack 1+ N \left (\frac{|A_N|}{\epsilon}\right
)^{1/N}\right\rbrack,\label{ae9}
\end{eqnarray}
which returns Eq. (\ref{ae3b}).

\section{The two-fluid model}
\label{sec_intro}

In this Appendix, we determine the two-fluid model corresponding to 
the GLDF with $A_i=A_n
\delta_{i,n}$. In particular, we establish the EoS of the DE
in the two-fluid model.

The GLDF is a one-fluid model (i.e. a UDM model) unifying DM and DE. The
pressure and the energy density are given by
\begin{eqnarray}
P=A_n \ln^n\left (\frac{\rho_m}{\rho_P}\right ),
\label{ae13}
\end{eqnarray}
\begin{eqnarray}
\epsilon=\rho_m c^2-A_n \frac{\rho_m}{\rho_P} \Gamma\left\lbrack n+1, \ln\left
(\frac{\rho_m}{\rho_P}\right )\right\rbrack=\rho_m
c^2+u=\epsilon_m+\epsilon_{de}.
\label{ae14}
\end{eqnarray}
Concerning the evolution of the homogeneous background, this one-fluid model
is equivalent\footnote{The equivalence between the one-fluid model and the
two-fluid model is lost when we consider the formation of structures.} to a
two-fluid model made of pressureless DM with an EoS $P_{m}=0$ 
giving $\epsilon_{m}=\Omega_{m,0}\epsilon_0/a^3$ and DE with an EoS
$P_{de}(\epsilon_{de})$ giving $\epsilon_{de}=u(a)$.  Noting that
$P=P_{m}+P_{de}=P_{de}$, the EoS $P_{de}(\epsilon_{de})$  of DE is determined in
parametric form by the equations
\begin{eqnarray}
P_{de}=A_n \ln^n\left (\frac{\rho_m}{\rho_P}\right ),
\label{ae15}
\end{eqnarray}
\begin{eqnarray}
\epsilon_{de}=-A_n \frac{\rho_m}{\rho_P} \Gamma\left\lbrack n+1,
\ln\left
(\frac{\rho_m}{\rho_P}\right )\right\rbrack.
\label{ae16}
\end{eqnarray}
Eliminating $\rho_m$ between these two expressions
we get
the EoS of DE under the reversed form $\epsilon_{de}(P_{de})$ as
\begin{eqnarray}
\epsilon_{de}=-A_n e^{\mp\left |\frac{P_{de}}{A_n}\right |^{1/n}}
\Gamma\left\lbrack
n+1,\mp\left |
\frac{P_{de}}{A_n}\right |^{1/n}\right\rbrack,
\label{ae19}
\end{eqnarray}
where the upper sign corresponds to the most relevant case $\rho_m<\rho_P$
and
the lower sign corresponds to $\rho_m>\rho_P$.

Let us check that the two-fluid model returns the results of the one-fluid
model for the homogeneous background. The evolution $\epsilon_{de}(a)$ of the DE
density with the scale factor can be obtained from the energy
conservation equation
\begin{equation}
\frac{d\epsilon_{de}}{da}+\frac{3}{a}(\epsilon_{de}+P_{de})=0 \qquad
\Leftrightarrow \qquad \int \frac{d\epsilon_{de}}{\epsilon_{de}+P_{de}}=-3\ln a
\label{ae19b}
\end{equation}
with the EoS $P_{de}(\epsilon_{de})$ defined by Eqs.
(\ref{ae15}) and (\ref{ae16}). At this stage, $\rho_m$ is just a dummy
variable. It is easy to establish that
\begin{eqnarray}
\epsilon_{de}'(\rho_m)=-A_n n \frac{\rho_P}{\rho_m^2} \Gamma\left\lbrack n,
\ln\left
(\frac{\rho_m}{\rho_P}\right )\right\rbrack
\label{ae16b}
\end{eqnarray}
and 
\begin{eqnarray}
P_{de}(\rho_m)+\epsilon_{de}(\rho_m)=-A_n n \frac{\rho_P}{\rho_m}
\Gamma\left\lbrack n,
\ln\left
(\frac{\rho_m}{\rho_P}\right )\right\rbrack.
\label{ae16c}
\end{eqnarray}
Therefore, we have the identity
\begin{eqnarray}
P_{de}(\rho_m)+\epsilon_{de}(\rho_m)=\rho_m \epsilon_{de}'(\rho_m),
\label{ae16d}
\end{eqnarray}
and the energy
conservation equation (\ref{ae19b}) becomes
\begin{equation}
\frac{d\rho_{m}}{da}+\frac{3}{a}\rho_{m}=0\qquad \Leftrightarrow
\qquad \int
\frac{d\rho_{m}}{\rho_m}=-3\ln a,
\label{ae19c}
\end{equation}
implying that $\rho_m\propto a^{-3}$. This leads to results consistent with
those of Sec. \ref{sec_part}. However, we cannot establish that $\rho_m$ is the
rest-mass (or DM) density (they could differ by a multiplicative constant). This
implies that we cannot determine $A_n$ in the two-fluid model contrary to the
one-fluid model.  More explicit results are given below.

\subsection{The case $n=1$}

For $n=1$, Eq. (\ref{ae19}) reduces to  the affine EoS
\begin{eqnarray}
P_{de}=-\epsilon_{de}-A
\label{ae20}
\end{eqnarray}
with $A>0$. This is the EoS of DE
corresponding to the original logotropic gas \cite{chavanis2017}. It coincides
with
the asymptotic EoS (\ref{ae10}) of the LDF seen as a one-fluid  (UDM) model. 
The affine EoS (\ref{ae20}) was first introduced and studied in
\cite{cosmopoly3}.

Let us
check that the two-fluid model returns the results of the one-fluid model.
Integrating the  energy conservation equation (\ref{ae19b}) for DE with the
EoS (\ref{ae20}) we
obtain
\begin{eqnarray}
\epsilon_{de}=3A\ln\left (\frac{a}{a_*}\right ),
\label{ae22}
\end{eqnarray}
where $a_*$ is a constant of integration. If we add the contribution of
pressureless DM, we find that the total energy density is given by
\begin{eqnarray}
\epsilon=\frac{\Omega_{m,0}\epsilon_0}{a^3}+3A\ln\left (\frac{a}{a_*}\right
).
\label{ae23}
\end{eqnarray}
Using the condition (at $a=1$)
\begin{eqnarray}
\epsilon_0(1-\Omega_{m,0})=-3A\ln a_*,
\label{ae24}
\end{eqnarray}
we can rewrite Eq. (\ref{ae23}) under the form
\begin{eqnarray}
\epsilon=\frac{\Omega_{m,0}\epsilon_0}{a^3}+3A\ln
a+\epsilon_0(1-\Omega_{m,0}).
\label{ae25}
\end{eqnarray}
This expression is consistent with Eq. (\ref{eq27}) of the one-fluid model if
we set $A=B\epsilon_0(1-\Omega_{m,0})$,
except that the two-fluid model does {\it not} determine the value of the
constant $A$, contrary to the one-fluid model. Indeed, the Planck scale
$\rho_P$ does not occur in Eq. (\ref{ae20}), unlike Eq. (\ref{eq1bo}). This is a
huge advantage of the
one-fluid model \cite{chavanis2015}.

\subsection{The case $n=2$}

For $n=2$, Eq. (\ref{ae19}) reduces to 
\begin{eqnarray}
\epsilon_{de}=-2A_2\left
(1\mp\sqrt{\frac{P_{de}}{A_2}}+\frac{P_{de}}{2A_2}\right )
\label{ae26}
\end{eqnarray}
with $A_2<0$. This yields a second degree equation for $\sqrt{P_{de}/A_2}$
of the form 
\begin{eqnarray}
\frac{P_{de}}{A_2}\mp
2\sqrt{\frac{P_{de}}{A_2}}+2+\frac{\epsilon_{de}}{A_2}=0.
\label{ae27}
\end{eqnarray}
When $\rho_m>\rho_P$, Eq. (\ref{ae27}) with the lower sign has just one solution
\begin{eqnarray}
\sqrt{-\frac{P_{de}}{|A_2|}}=-1+\sqrt{\frac{\epsilon_{de}}{
|A_2|}-1}.
\end{eqnarray}
As $\rho_m$ decreases from $+\infty$ to $\rho_P$, the DE
density
$\epsilon_{de}$ decreases from $+\infty$ to $2|A_2|$ and the pressure $P_{de}$
increases from $-\infty$ to $0$ (see Sec. \ref{sec_casd}). This leads to the
EoS
\begin{eqnarray}
P_{de}=-|A_2| \left (-1+\sqrt{\frac{\epsilon_{de}}{|A_2|}-1}\right )^2
\end{eqnarray}
or, equivalently,
\begin{eqnarray}
P_{de}=-\epsilon_{de}+2|A_2|\sqrt{\frac{\epsilon_{de}}{|A_2|}-1},
\label{mar1}
\end{eqnarray}
which is valid for $\epsilon_{de}\ge 2|A_2|$. For
$\epsilon_{de}\rightarrow +\infty$, Eq. (\ref{mar1}) reduces to 
\begin{eqnarray}
P_{de}\simeq -\epsilon_{de}+2|A_2|^{1/2} \sqrt{\epsilon_{de}}.
\end{eqnarray}
When $\rho_m<\rho_P$, the solutions of Eq. (\ref{ae27}) with the upper sign are
\begin{eqnarray}
\sqrt{-\frac{P_{de}}{|A_2|}}=1\pm\sqrt{\frac{\epsilon_{de}}{|A_2|}-1}.
\label{ae28}
\end{eqnarray}
As $\rho_m$ goes from $\rho_P$ to $0$, the DE density
$\epsilon_{de}$ first decreases from $2|A_2|$ to  $|A_2|$ then increases to
$+\infty$ while the pressure $P_{de}$ decreases from $0$ to $-\infty$ (see Sec.
\ref{sec_casd}).  Equation (\ref{ae27}) has
two solutions for $|A_2|\le \epsilon_{de}\le 2|A_2|$, one solution (with the
upper sign) for $\epsilon_{de}\ge
2|A_2|$ and no solution
for $\epsilon_{de}\le |A_2|$.
This leads to the EoS
\begin{eqnarray}
P_{de}=-|A_2| \left (1\pm\sqrt{\frac{\epsilon_{de}}{|A_2|}-1}\right )^2
\label{ae29}
\end{eqnarray}
or, equivalently,
\begin{eqnarray}
P_{de}=-\epsilon_{de}\mp 2|A_2| \sqrt{\frac{\epsilon_{de}}{|A_2|}-1},
\label{ae30}
\end{eqnarray}
which is valid with the two signs for $|A_2|\le \epsilon_{de}\le 2|A_2|$ and
with the upper sign for $\epsilon_{de}\ge 2|A_2|$. This is
the EoS of DE corresponding to the GLDF in the case $n=2$. For
$\epsilon_{de}\rightarrow +\infty$, Eq. (\ref{ae30}) reduces to 
\begin{eqnarray}
P_{de}\simeq -\epsilon_{de}- 2|A_2|^{1/2} \sqrt{\epsilon_{de}},
\label{ae31}
\end{eqnarray}
which coincides with the asymptotic EoS (\ref{ae12}) of the GLDF seen as a
one-fluid (UDM) model.

Let us
check that the two-fluid model returns the results of the one-fluid model.
Integrating the  energy conservation equation (\ref{ae19b}) for DE with the
EoS (\ref{mar1}) or (\ref{ae30}) we
obtain 
\begin{eqnarray}
\frac{\epsilon_{de}}{|A_2|}=1+9\ln^2\left (\frac{a}{a_*}\right ),
\label{ae34}
\end{eqnarray}
where $a_*$ is a constant of integration. 
If we add the contribution of
pressureless DM, we find that the total energy density is given by
\begin{eqnarray}
\epsilon=\frac{\Omega_{m,0}\epsilon_0}{a^3}+|A_2|\left\lbrack 1+9\ln^2\left
(\frac{a}{a_*}\right )\right\rbrack.
\label{ae35}
\end{eqnarray}
Using the condition (at $a=1$)
\begin{eqnarray}
\epsilon_0(1-\Omega_{m,0})=|A_2|\left ( 1+9\ln^2a_*\right ),
\label{ae36}
\end{eqnarray}
we can rewrite Eq. (\ref{ae35}) under the form
\begin{eqnarray}
\epsilon=\frac{\Omega_{m,0}\epsilon_0}{a^3}+\epsilon_0(1-\Omega_{m,0}
)\left\lbrack 1+\frac{9|A_2|\ln^2
a}{\epsilon_0(1-\Omega_{m,0})}+\frac{6|A_2|\ln
a}{\epsilon_0(1-\Omega_{m,0})}\sqrt{\frac{\epsilon_0(1-\Omega_{m,0})}{|A_2|}
-1}\right\rbrack.
\end{eqnarray}
This expression is
consistent with Eq.
(\ref{eq29b}) of the one-fluid model if we set
$|A_2|/[\epsilon_0(1-\Omega_{m,0})]=B^2/(1+B^2)$, except that the two-fluid
model does {\it not} determine the value of the
constant $A_2$, contrary to the one-fluid model. This is a huge advantage of the
one-fluid model \cite{chavanis2015}. 

\section{Present proportion of dark matter and dark energy}
\label{sec_arg}

In this Appendix we recall the argument of \cite{chavanis2018} leading to a
prediction of the  present proportion of DM and DE in the universe and take into
account the
presence of baryons.

The original logotropic
model \cite{chavanis2015,chavanis2016,chavanis2017,chavanis2018} is based on
the EoS
\begin{eqnarray}
P  =    A \ln \left(\frac{\rho_{dm}}{\rho_{P}} \right),
\label{f1}
\end{eqnarray}
where $\rho_{dm}$ is the rest-mass density of the LDF, $A$ is a new fundamental
constant
of physics (superseding the cosmological constant) and $\rho_P$ is the Planck
density. This EoS provides a unification of DM and DE. It is very interesting
that the Planck density appears in this EoS in order to make the variable in the
logarithm dimensionless. This implies that quantum effects play a certain role
in the late universe where the logotropic model applies. The
rest-mass density evolves as [see Eq. (\ref{eq15})]
\begin{eqnarray}
\rho_{dm}=\frac{\Omega_{dm,0}\epsilon_0/c^2}{a^3},
\label{f2}
\end{eqnarray}
and it plays the role of DM. As a result $\Omega_{dm,0}$ is interpreted as the
present proportion of DM in the universe. The
energy density of the LDF is [see Eq. (\ref{nrj})]
\begin{eqnarray}
\epsilon_{df}=\rho_{dm} c^2-A\left\lbrack 1+\ln\left
(\frac{\rho_{dm}}{\rho_P}\right
)\right\rbrack=\epsilon_{dm}+\epsilon_{de},
\label{f3}
\end{eqnarray}
where the first term (rest-mass)  is interpreted as DM and the second term
(internal energy) as DE. We must also include the contribution of baryons which
form a pressureless gas ($P_b=0$). Their energy density evolves as
\begin{eqnarray}
\epsilon_b=\frac{\Omega_{b,0}\epsilon_0}{a^3},
\label{f4}
\end{eqnarray}
where $\Omega_{b,0}$ is the present proportion of baryons in the universe. The
total energy density is therefore
\begin{eqnarray}
\epsilon=\rho_{dm} c^2-A\left\lbrack 1+\ln\left (\frac{\rho_{dm}}{\rho_P}\right
)\right\rbrack+\epsilon_{b}.
\label{f5}
\end{eqnarray}
Substituting Eq. (\ref{f2}) into Eq. (\ref{f5}), we get
\begin{eqnarray}
\epsilon=\frac{\Omega_{dm,0}\epsilon_0}{a^3}-A\left\lbrack 1+\ln\left
(\frac{\Omega_{dm,0}\epsilon_0}{\rho_P c^2 a^3}\right
)\right\rbrack+\frac{\Omega_{b,0}\epsilon_0}{a^3}.
\label{f6}
\end{eqnarray}
Applying this relation at the present time ($a=1$) and introducing
the present proportion of DE in the
universe $\Omega_{de,0}=1-\Omega_{dm,0}-\Omega_{b,0}$ we
get
\begin{eqnarray}
A=\frac{\Omega_{de,0}\epsilon_0}{
\ln\left
(\frac{\rho_P c^2}{\Omega_{dm,0}\epsilon_0}\right
)-1}.
\label{f7}
\end{eqnarray}
Introducing the present DE density $\rho_{\Lambda}\equiv
\Omega_{de,0}\epsilon_0/c^2$, we can rewrite the foregoing equation as
\begin{eqnarray}
A=\frac{\rho_{\Lambda}c^2}{
\ln\left
(\frac{\rho_P}{\rho_{\Lambda}}\right )+\ln\left (
\frac{\Omega_{de,0}}{1-\Omega_{de,0}-\Omega_{b,0}}\right
)-1}.
\label{f8}
\end{eqnarray}
We now postulate that \cite{chavanis2018} 
\begin{eqnarray}
A=\frac{\rho_{\Lambda}c^2}{
\ln\left
(\frac{\rho_P}{\rho_{\Lambda}}\right )}
\label{f9}
\end{eqnarray}
or, equivalently, that $B=\ln ({\rho_P}/{\rho_{\Lambda}})$. This
implies 
\begin{eqnarray}
\frac{\Omega_{de,0}}{1-\Omega_{de,0}-\Omega_{b,0}}=e,
\label{f10}
\end{eqnarray}
determining the present proportion of DM and DE  \cite{chavanis2018} 
\begin{eqnarray}
\Omega_{de,0}^{\rm th}=\frac{e}{1+e}(1-\Omega_{b,0}), \qquad \Omega_{dm,0}^{\rm
th}=\frac{1}{1+e}(1-\Omega_{b,0}).
\label{f11}
\end{eqnarray}
If we neglect baryonic matter $\Omega_{b,0}=0$ we obtain the pure numbers
$\Omega_{de,0}^{\rm th}=\frac{e}{1+e}=0.731059...$ and $\Omega_{dm,0}^{\rm
th}=\frac{1}{1+e}=0.268941...$ which give the correct
proportions $70\%$ and $25\%$ of DE and DM \cite{chavanis2018}. If we take 
baryonic matter into account and use the measured value of
$\Omega_{b,0}=0.0486\pm 0.0010$, we get
$\Omega_{de,0}^{\rm th}=0.6955\pm 0.0007$ and $\Omega_{dm,0}^{\rm th}=0.2559\pm
0.0003$ which are very close to the
observed values $\Omega_{de,0}=0.6911\pm 0.0062$  and $\Omega_{dm,0}=0.2589\pm
0.0057$ within the error bars. The postulate from Eq.
(\ref{f9}) means that the fundamental constant $A$ is equal to the {\it present}
DE energy density (more precisely
$\rho_{\Lambda}c^2/\ln ({\rho_P}/{\rho_{\Lambda}})$). This can be viewed as
a strong cosmic coincidence \cite{chavanis2018} giving to our epoch a central
place in the history of the universe.

\end{document}